\documentclass{article}
\usepackage{amsmath,amssymb,amsthm,mathrsfs,bm,hyperref}

\usepackage{ae,aecompl}

\title{Lorentz Covariant \(\kappa\)-Minkowski Spacetime}

\author{Ludwik D\k{a}browski,
Micha\l\ Godli\'nski and 	
Gherardo Piacitelli\\
\footnotesize SISSA, Via Bonomea 265, 34136, Trieste, Italy}
\begin{document}
\maketitle

\begin{abstract}
In recent years, different views on the interpretation of Lorentz covariance
of non commuting coordinates have been discussed. By a general procedure, we 
construct the minimal canonical central covariantisation of the 
\(\kappa\)-Minkowski spacetime. Here, undeformed Lorentz covariance is 
implemented by unitary operators, in the presence of two dimensionful 
parameters. We then show that, though the usual
\(\kappa\)-Minkowski spacetime is covariant under 
deformed (or twisted) Lorentz action, the resulting framework is equivalent
to taking a non  covariant restriction of the
covariantised model.  We conclude
with some general comments on the approach of deformed covariance. 
\end{abstract}

\section{Introduction}
The \(\kappa\)-Minkowski spacetime is defined by the commutation relations
\begin{subequations}
\label{eq:kMink}
\begin{gather}
\label{eq:kMink_time}
[X^0,X^j]=i X^j,\quad j=1,2,3,\\
\label{eq:kMink_space}
[X^j,X^k]=0,\quad j,k=1,2,3, 
\end{gather}
\end{subequations}
among the four selfadjoint operators 
\[
(X^\mu)=(X^0,X^1,X^2,X^3),
\]
to be interpreted as the coordinates of a noncommutative version of the
usual Minkowski spacetime \cite{Lukierski:1991pn,Majid:1994cy}; 
see also \cite{Dabrowski:2010yk,Dabrowski:2009hv} for
a detailed discussion. The name refers to the traditional notation
\(\kappa\) for the inverse length scale; 
here we take natural units where \(\kappa=1\). 

We do not consider commutation
relations among coordinates and momenta, since we do not aim to a 
``more noncommutative'' version of quantum mechanics. Indeed, energies involved
at the Planck scale are so high that their description is expected to require
a generalisation of quantum field theory. As an intermediate step, we are 
interested in generalizing the position space, not the quantum mechanical 
phase space. This should give a noncommutative replacement of the pointwise product
of quantum fields, for a flat model (few processes of very high energy; see \cite{Doplicher:1994tu} for a discussion).

It was observed (see \cite{Lukierski:2002ii} and references therein), 
that the above relations can be generalised in the form
\begin{equation}\label{eq:lukierski}
[X^\mu, X^\nu]=i(v^\mu X^\nu-v^\nu X^\mu),
\end{equation}
where \(v\) is a fixed 4-vector in \(\mathbb R^4\). The standard timelike
choice  \(v_{(0)}=(1,0,0,0)\) reproduces precisely the usual \(\kappa\)-Minkowski
spacetime \eqref{eq:kMink}.

In this model,
covariance under Lorentz boosts and space translations has been sacrificed
from the beginning, and replaced by \(\kappa\)-Poincar\'e symmetry, in the
framework of quantum groups. The model has been under extensive investigation
for almost 20 years (and still is), because of its nice 
mathematical features which make it a convenient framework where to test 
general ideas. See e.g.\ \cite{KowalskiGlikman:2004qa} 
for a recent review, focused on therecently proposed 
connections with Doubly Special Relativity. 

From the point of view of spacetime quantisation, 
it has some inconvenient features:
(i) the uncertainty relations are not sufficient to prevent
an exceedingly high energy transfer to the geometric background by 
localisation, at least very close to the centre of space where the model is ``nearly commutative'', (ii) on the 
contrary, at large scale, the model quickly becomes non commutative, so that
e.g.\ it is not clear how to consistently formulate LHC physics 
already as far as 1mm from the centre of the \(\kappa\)-Minkowski, 
if \(1/\kappa\) is of order
of Planck length (see \cite{Dabrowski:2010yk} for this and other estimates).

Here,  
we construct a new model which is the smallest extension of 
the \(\kappa\)-Minkowski spacetime, among those enjoying Lorentz covariance 
in the sense of Wigner. 
For this reason
we call this model the {\itshape Lorentz 
covariant \(\kappa\)-Minkowski spacetime}. It is defined by the relations
\begin{subequations}
\label{eq:kappa_cov_rel}
\begin{gather}
\label{eq:kappa_cov_rel_xx}
[X^\mu, X^\nu]=i(V^\mu X^\nu-V^\nu X^\mu),\\
\label{eq:kappa_cov_rel_xv}
[X^\mu,V^\nu]=0,\\
\label{eq:kappa_cov_rel_vv}
[V^\mu,V^\nu]=0,\\
\label{eq:kappa_cov_rel_mass_shell}
V_\mu V^\mu=I, 
\end{gather}
\end{subequations}
where \(I\) is identity operator and \(V\) is a purely vectorial
quantity. If we took \(V\) to have the form 
\(V^\mu=v^\mu I\) for some ordinary 4-vector \(v\), we would
fall back to
timelike models in the class considered in \cite{Lukierski:2002ii}.  
In the spirit of \cite{Doplicher:1994tu}, we propose instead to 
allow each \(V^\mu\) to be a
non trivial operator (not a multiple of the identity). 
In this case, the model admits a fully Lorentz
covariant representation, namely such that there is a strongly 
continuous unitary representation \(U\) of the Lorentz group 
\(\mathscr L=O(1,3)\), fulfilling
\begin{subequations}
\label{eq:kappa_cov_unit}
\begin{align}
U(\varLambda^{-1})X^\mu U(\varLambda)&={\varLambda^\mu}_\nu X^\nu,\\
U(\varLambda^{-1})V^\mu U(\varLambda)&={\varLambda^\mu}_\nu V^\nu,
\end{align}
\end{subequations}
which imply
\begin{equation}\label{eq:super_explicit}
U(\varLambda^{-1})[X^\mu,X^\nu] U(\varLambda)=i
{\varLambda^\mu}_{\mu'} {\varLambda^\nu}_{\nu'}
(V^{\mu'} X^{\nu'}-V^{\nu'} X^{\mu'}). 
\end{equation}
Note that this is stronger than requiring simple form--covariance. For example, 
\eqref{eq:super_explicit} cannot be obtained 
if the components of \(V\) are ordinary numbers, namely 
\(V^\mu=v^\mu I\) as in \eqref{eq:lukierski}. Indeed, 
unitary operators commute with  multiples of the identity, so that 
\(
U(\varLambda^{-1})v^\mu X^\nu U(\varLambda)=
v^\mu {\varLambda^\nu}_{\nu'} X^{\nu'}
\)
(no action on the index \(\mu\)). For a covariant representation in the sense
of \eqref{eq:kappa_cov_unit},
the (generalised) common eigenvalues \(v^\mu\) of the pairwise commuting 
operators \(V^\mu\) describe precisely the two--sheeted mass 1 hyperboloid.
 
Since star--products of symbols are a shadow of operator products,
{\itshape full covariance} is precisely the appropriate notion 
for discussing the behaviour of star--products under Lorentz 
transformations of the symbols, in the spirit of Wigner theorem on quantum 
symmetries.

Note also that this provides yet another example 
(the first one being the DFR model
\cite{Doplicher:1994tu}) where Lorentz symmetry is meaningful
in the presence of an invariant characteristic length  scale in a
noncommutative setting, without deforming the Lorentz group nor its action. 
Even more,
our covariantisation procedure gives a tool for constructing a rich family
of such examples, where there are two invariant dimensionful parameters:
the light speed \(c\) (here also set to one) and the inverse Planck length 
\(\kappa\). In other 
words, a doubly special relativity framework does not require necessarily a 
deformation of the Lorentz group.

Obeying to the motto ``no deformation without representation'',
the covariant representation is explicitly described in the Mathematical 
Appendix,
where the C*-algebra of the model also is discussed.

The covariant \(\kappa\)-Minkowski spacetime 
arises as the {\itshape minimal canonical 
central covariantisation} of the usual
\(\kappa\)-Minkowski model, presented in section
\ref{sec:be_wise}. Our recipe for central 
covariantisation is minimal, in the sense that the family of 
(equivalence classes of) irreducible
representations cannot be pruned any further
without destroying
the possibility of building up a covariant representation.

It is clear that the \(\kappa\)-Minkowski relations \eqref{eq:kMink}
can be obtained by supplementing relations \eqref{eq:kappa_cov_rel}
with the constraint 
\begin{equation}\label{eq:constraint}
V^0=I,
\end{equation}
which, by (\ref{eq:kappa_cov_rel_mass_shell},\ref{eq:kappa_cov_rel_vv}), 
entails \(V^1=V^2=V^3=0\).
Condition \eqref{eq:constraint} may be regarded as a criterion 
for selecting the usual \(\kappa\)-Minkowski out of the fully covariant model,
as a subrepresentation of the latter. 
In this sense the 
\(\kappa\)-Minkowski model is a reduction of the fully covariant model,
which reproduces the very same relationship \cite{Piacitelli:2009fa}
between the full DFR model
\cite{Doplicher:1994tu} and  the
reduced DFR model (a member of the family of the so called ``canonical 
quantum spacetimes'').

Since the class of representations (or equivalently of localisation states, 
according to the GNS theorem) selected by \eqref{eq:constraint}
is not invariant, the resulting theory,
though formally covariant, breaks the relativity principle, in the sense 
already discussed in \cite{Piacitelli:2009tb}: the selection of admissible
representations and states is defined with respect to 
some preferred observer in his/her special
Lorentz frame. 

To complete the picture, we show that \(\kappa\)-Minkowski spacetime admits
an approach based on ``deformed 
covariance'', which parallels the ``twisted covariance'' of
\cite{Chaichian:2004za,Wess:2003da,Chaichian:2004yh}. 
This provides a deformation operator 
(``twist'') realising the star product, in the case of
timelike choices of \(v\) in \eqref{eq:lukierski}
(for a discussion of the lightlike case in relation with the classical
r-matrices, see \cite{Lukierski:2002ii}). 
However, the resulting
formalism is found plainly 
equivalent to rejecting all the representations with \(V^0\neq I\). 

We finally draw some conclusions, in particular concerning the r\^ole of 
``twists'' in spacetime quantisation,
and the (lack of) physical motivations for rejecting otherwise admissible
localisation states.

\section{Minimal Canonical Central Covariantisation}
\label{sec:be_wise}

Assume that we have a finitely generated 
model of a flat, noncommutative spacetime described in terms of the selfadjoint operators \(X^N\), \(N=0,1,2\dotsc \bar N\), where \(3\leqslant\bar N<\infty\). The first four operators
\(X^0,\dotsc,X^3\) are to be interpreted as the non commutative coordinates
of a 4-event.

In order to get compact equations, 
we set \((g^{MN})=\text{diag}(1,-1,-1\dotsc,-1)\) as a 
\((\bar N+1)\times(\bar N+1)\) matrix, and we extend the usual conventions about implicit summation, 
raising and lowering to the full set of indices; this is a purely formal
convention, with no underlying interpretation, and no loss of generality. When
we use Greek symbols \(\mu,\nu,\dotsc\) we understand dummy indices 
running in the set \(\{0,1,2,3\}\); Latin dummy indices  \(m,n,\dotsc\) 
run in \(\{4,5,\dotsc,\bar N\}\), while capital Latin dummy indices
\(M,N,\dotsc\) run in the full set \(\{0,1,2,\dotsc,\bar N\}\). 

This done, we assume the commutation relations
to have the form
\begin{equation}\label{eq:initial_mod}
[X^M,X^N]=i{{\theta}^{MN}}_RX^R
\end{equation}
for a given real tensor \(\theta\), antisymmetric 
in the two upper indices, and fulfilling the constraints imposed
by the Jacobi identity; \(i\) is the imaginary unit.

The above framework incorporates a large class of models, among which we mention
\begin{enumerate} 
\item 
The canonical quantum spacetime, where
\(\bar N=4\), \(X^4=I\), and 
\begin{gather*}
{{\theta}^{M N}}_\rho=0,\quad
{{\theta}^{M n}}_4=0.
\end{gather*} 
If we choose \({\sigma_{(0)}}^{\mu\nu}={{\theta}^{\mu\nu}}_4\) to be the standard
symplectic matrix (or any of its Lorentz transforms), we obtain the reduced DFR model \cite{Doplicher:1994tu}.

\item 
The  \(\kappa\)-Minkowski spacetime, where \(\bar N=3\), there only are
Greek dummy indices, and 
\[
{{\theta}^{\mu\nu}}_\rho={g^\mu}_\rho{v_{(0)}}^\nu-{g^\nu}_\rho{v_{(0)}}^\mu
\]
with \({{v_{(0)}}}=(1,0,0,0)\).
\end{enumerate}

We now wish to consider the effect of Lorentz transformations on 
the relations among the physical coordinates. We take the point of 
view that the additional operators \(X^4,\dotsc,X^{\bar N}\)
correspond to inner degrees of freedom of the noncommutative spacetime
(a fixed background). Hence they are unaffected by a Lorentz transformation. 
It is clear that,
defining new operators \(X'\) by 
\begin{gather*}
{X'}^{\;\mu}={\varLambda^{\mu}}_{\nu}{X^{\nu}},\\
{X'}^{\;n}=X^n,
\end{gather*}
they fulfil
\[
[{X'}^{\;M},{X'}^{\;N}]=i{{\theta'}^{\;MN}}_R{X'}^{\;R},
\]
where \(\theta'\) is obtained by transforming the Lorentz (Greek) 
values only of the 
indices of \(\theta\). 

We may regard \(\theta\), describing the commutator of the initial 
non covariant model, as a collection of Lorentz tensors,
parametrized by the non Lorentz (lower case Latin) values of the indices. 
In particular, fixing two indices to take non Lorentz values, 
we have vectors with index \(\mu\), parametrised
by \(m,n\):
\[
{{\theta}^{\mu m}}_n,{{\theta}^{m \mu}}_n,{{\theta}^{mn}}_\mu;
\]
the second is redundant because of antisymmetry, so that we have 
the following independent Lorentz invariants, labeled by \(n,m\):
\begin{gather*}
\phi_1(\theta)={{\theta}^{\mu m}}_n{{{\theta}_{\mu}}^m}{}_n,\\
\phi_2(\theta)=\theta^{mn\mu}{\theta^{mn}}_\mu,\\
\text{(no summation over Latin indices)}.
\end{gather*}
With two Lorentz indices, we have 
\[
{{\theta}^{\mu\nu}}_n,\quad 
{{\theta}^{\mu n}}_\nu 
\]
where again we used antisymmetry in the upper index 
to discard the redundant one. For the first of them, we have
the following independent Lorentz invariants
\begin{gather*}
\phi_3(\theta)=
{{\theta}^{\mu\nu}}_n\theta_{\mu\nu n},\\
\phi_4(\theta)=({{\theta}^{\mu\nu}}_n{(\ast{{\theta}}_n)_{\mu\nu}})^2,\\
\text{(no summation over Latin indices)},
\end{gather*}
where \(\ast{{\theta}}_n\)
stands for the Hodge dual of \({{\theta}^{\mu\nu}}_n\) in the indices 
\(\mu,\nu\), holding \(n\) fixed. We will not list all the invariants
of \({{\theta}^{\mu n}}_\nu\) (which is not symmetric) and of the 
only tensor with three Lorentz indices \({{\theta}^{\mu \nu}}_\rho\),
not to burden the presentation, and because we will not need them here.

Hence, we may define a central covariantisation of the initial 
model by simply turning the numbers \({\theta^{MN}}_R\) into central
operators \({\Theta^{MN}}_R\), setting
\begin{subequations}
\label{eq:min_cov_ctrl}
\begin{align}
\label{eq:a}
[X^M,X^N]&=i{\Theta^{MN}}_RX^R,\\
[X^M,{\Theta^{MN}}_R]&=0,\\
\label{eq:c}
[{\Theta^{MN}}_R,{\Theta^{M'N'}}_{R'}]&=0.
\end{align}
The above certainly contains the given initial model as a subrepresentation.
However, this would add way too many new representations to the initial model; 
the only remaining dependence on the initial model
is on the number of generators, and in a sense we may regard it as the maximal
central covariantisation. 
We want a covariant model that fulfils two 
natural requirements, namely, (i) it contains the initial model as a 
subrepresentation, and (ii) it is the smallest possible covariant extension
of the initial model. We obtain this by adding
all invariant constraints on the admissible representations, 
which are fulfilled by the initial model:
\[
\phi_j(\Theta)=\phi_j(\theta)I,\quad j=1,2,3,\dotsc;
\]
more explicitly,
\begin{gather}
{{\Theta}^{\mu m}}_n{{{\Theta}_{\mu}}^m}{}_n=
{{\theta}^{\mu m}}_n{{{\theta}_{\mu}}^m}{}_nI,\\
\Theta^{mn\mu}{\Theta^{mn}}_\mu=
\theta^{mn\mu}{\theta^{mn}}_\mu I,\\
{{\Theta}^{\mu\nu}}_n\Theta_{\mu\nu n}=
{{\theta}^{\mu\nu}}_n\theta_{\mu\nu n}I,\\
({{\Theta}^{\mu\nu}}_n{(\ast{{\Theta}}_n)_{\mu\nu}})^2=
({{\theta}^{\mu\nu}}_n{(\ast{{\theta}}_n)_{\mu\nu}})^2I,\\
\dots\\
\nonumber\text{(no summation over Latin indices)},
\end{gather}
\end{subequations}
where the ellipsis stand for the analogous equations corresponding
to the invariants which we did not write down explicitly.

Altogether, the relations \eqref{eq:min_cov_ctrl} define the minimal central 
covariantization of the relations \eqref{eq:initial_mod}. 
The above strategy is canonical, in the sense that it does not depend on any 
additional arbitrary choice, once the choice of the initial relations 
\eqref{eq:initial_mod} is made.

Let us see two applications of the above mentioned general strategy.
\begin{enumerate}
\item In the reduced DFR model, 
\(\theta\) contains the following Lorentz tensors, 
\begin{gather*}
{\theta^{\mu 4}}_4,\quad \text{(one Lorentz index)},\\
{\theta^{\mu\nu}}_4,\quad {\theta^{\mu 4}}_\nu \quad \text{(two Lorentz indices)},\\
{\theta^{\mu\nu}}_\rho \quad\text{(three Lorentz indices)}.
\end{gather*}
The only one which is not identically vanishing is
\({\theta^{\mu\nu}}_4={\sigma_{(0)}}^{\mu\nu}\). Two independent invariants of
this tensor are
\[
{\sigma_{(0)}}_{\mu\nu}{\sigma_{(0)}}^{\mu\nu}=0,
\quad
\frac 1{16}({\sigma_{(0)}}_{\mu\nu}(\ast\sigma_{(0)})^{\mu\nu})^2=1.
\]
Hence the covariantized model is defined by the Eqs. (\ref{eq:a}-\ref{eq:c}), 
complemented with the constraints
\begin{align*}
{\Theta^{\mu 4}}_N&=0,\\
{\Theta^{\mu \nu}}_\rho&=0,\\
{\Theta_{\mu\nu}}_4{\Theta^{\mu\nu}}_4&=0,\\
\left(\frac 14 {\Theta_{\mu\nu}}_4(\ast \Theta_4)^{\mu\nu}\right)^2&=X^4.
\end{align*}
With the identifications
\[
q^\mu=X^\mu,\quad Q^{\mu\nu}={\Theta^{\mu\nu}}_4,\quad I=X^4,
\]
we recognise the commutation relations of the DFR 
model~\cite{Doplicher:1994tu}.

\item 
The  \(\kappa\)-Minkowski spacetime, where \(\bar N=3\), dummy indices take 
only Greek values, and 
\[
{\theta^{\mu\nu}}_\rho={g^\mu}_\rho{v_{(0)}}^\nu-{g^\nu}_\rho{v_{(0)}}^\mu
\]
with \({{v_{(0)}}}=(1,0,0,0)\). In this case the tensor \(\theta\) has no 
traceless component 
and there only is the vectorial part. The tensor 
\(\theta\) is uniquely associated to \({v_{(0)}}\), 
since \({v_{(0)}}^\mu=(1/3){\theta^{\rho\mu}}_\rho\). So we may forget 
\(\theta\) and make our statements directly in terms of \({v_{(0)}}\),
whose only invariant is \(\phi({v_{(0)}})={v_{(0)}}_\mu {v_{(0)}}^\mu=1\). 
Then we turn \({v_{(0)}}\) into a vector \(V\) with operator entries.
A centrally covariantized model (not the maximal one, since 
tensors with traceless components are implicitly ruled out) is
given by (\ref{eq:kappa_cov_rel_xx},\ref{eq:kappa_cov_rel_xv},\ref{eq:kappa_cov_rel_vv}).
We restrict to the minimal covariantization through
the relation \(\phi(V)=\phi({v_{(0)}})I,\) namely~\eqref{eq:kappa_cov_rel_mass_shell}.

We show in the Appendix that there exists a covariant representation: 
selfadjoint operators \(X^\mu,V^\mu\) and a unitary representation \(U\)
of the Lorentz group, fulfilling \eqref{eq:kappa_cov_rel} and
\begin{gather*}
U(\varLambda^{-1})X^\mu U(\varLambda)={\varLambda^\mu}_\nu X^\nu,\\
U(\varLambda^{-1})V^\mu U(\varLambda)={\varLambda^\mu}_\nu V^\nu.
\end{gather*}

\end{enumerate}

\section{Deformed Covariance}
\label{sec:tw_cov} 
Another, apparently unrelated, approach to covariance is described
in this section; it was proposed independently by \cite{Chaichian:2004za,Wess:2003da},
to obtain quantum group deformations of the Lorentz group, in the case 
of the canonical quantum spacetime (see also \cite{Aschieri:2009zz}). 

To a certain extent, the basic argument is of considerable generality, and we 
will apply it to the \(\kappa\)-Minkowski case.

Let
\[
m(f\otimes g)=fg
\]
be the usual (commutative) pointwise multiplications of functions. Let
\[
(\alpha_\varLambda f)(x)=f(\varLambda^{-1}x)
\] 
be the usual action
of the Lorentz group, and \(\alpha^{(2)}_\varLambda=\alpha_\varLambda\otimes
\alpha_\varLambda\) so that
\[
\alpha^{(2)}_\varLambda(f\otimes g)(x,y)=f(\varLambda^{-1}x)g(\varLambda^{-1}y).
\]
The usual multiplication \(m\) intertwines the actions 
\(\alpha_\varLambda\) and \(\alpha^{(2)}_\varLambda\):
\begin{equation}\label{eq:ord_prod_cov}
m\circ\alpha^{(2)}_\varLambda=\alpha_\varLambda\circ m.
\end{equation}

Now, assume that an associative product \(\star_0\) of ordinary functions of 
\(\mathbb R^4\) is given.
Assume also that we are able to find an {\itshape invertible}
operator \(F\) such that
\begin{equation}
f\star_0 g=m\circ F(f\otimes g).
\end{equation}
Once such an \(F\) 
is found to exist, one may deform \(\alpha^{(2)}_\varLambda\) into
\[
\tilde\alpha^{(2)}_\varLambda:=F^{-1}
\circ (\alpha_\varLambda\otimes\alpha_\varLambda)\circ F.
\]
Let us now introduce the notation \(m_{\star_0}(f\otimes g)=f\star_0 g\).
We observe that
\[
m_{\star_0}\circ\tilde\alpha^{(2)}_\varLambda=
m\circ \alpha^{(2)}_\varLambda\circ F,
\]
while 
\[
\alpha_\varLambda\circ m_{\star_0}=\alpha_\varLambda\circ m\circ F,
\]
Since \(F\) is invertible, equality between the right hand sides of the two equations 
above is equivalent to \eqref{eq:ord_prod_cov}. It follows that the left hand sides 
of the above are equal, and \(m_{\star_0}\) fulfils ``deformed covariance'':
\begin{equation}\label{eq:def_cov}
m_{\star_0}\circ\tilde\alpha^{(2)}_\varLambda=\alpha_\varLambda\circ m_{\star_0},
\end{equation}
which is a deformation of  \eqref{eq:ord_prod_cov}.

This trick works whenever an operator such as \(F\) exists. Its precise form
is not relevant in the game, and actually when it exists it 
is highly non unique. \(F\) is called a twist in the case of the canonical
quantum spacetime, since it reproduces (in momentum space) the twisted 
convolution product according to the terminology of \cite{vonNeumann:1931}.

We may apply these ideas to the case of the \(\kappa\)-Minkowski 
spacetime as well, instead of covariantizing it as described in the
preceding sections. 
We keep up with the original, non covariant coordinates \(X^\mu\), 
fulfilling \eqref{eq:kMink} (as operators on some Hilbert space \(\mathfrak H\)). Under the quantization 
prescription {\itshape \`a la Weyl}
\begin{equation}\label{eq:weyl_quant}
f(X)=\frac 1{(2\pi)^2}\int d\alpha\;\hat f(\alpha) e^{i\alpha_\mu X^\mu}
\end{equation}
(as an operator on \(\mathfrak H\)), 
we define a star product through
\begin{equation}\label{eq:star_prod}
(f\star_0 g)(X)=f(X)g(X),
\end{equation}
where \(f,g\) are ordinary functions of \(\mathbb R^4\), often called 
(Weyl) symbols 
in the theory of pseudodifferential operators.
If properly treated, the symbolic calculus can be used as an equivalent 
replacement of the underlying operator algebra. 

We finally prove the existence of a suitable 
deformation operator \(F\)
in the case of \(\kappa\)-Minkowski spacetime. We refrain from giving a 
complete computation, since the functional form of \(F\) is irrelevant.

Note that  
we prefer to call \(F\) the {\itshape deformation operator}, since 
in this case it is not associated with a twisted
convolution: indeed, contrary to the case of canonical commutation relations,
the appropriate class of Weyl operators 
is closed under operator products \cite{Dabrowski:2009hv}.
The relations
with twists in the sense of Drinfel'd and the Hopf--theoretical approach 
will be discussed elsewhere.

Both in the present case and in the case of the twisted product, it is
easier to work in Fourier space and look for an invertible operator
\(T\) such that
\begin{equation}
\label{eq:tw_four}
(c\circ T)(\hat f\!\otimes\!\hat g)=\hat f\ast_0\hat g,
\end{equation}
where \(\ast_0\) is deformed product in Fourier space, and \(c\)
is ordinary convolution. 
Then the desired \(F\) is obtained by
\[
\widehat{F f\!\otimes\! g}=T(\hat f\!\otimes\!\hat g).
\] 
Moreover, since the algebra of relations in \(3+1\) dimensions is a central
extension of the algebra in \(1+1\) dimensions (\cite{Dabrowski:2010yk}; see 
Mathematical Appendix, eqn.s \eqref{eq:radial_rels}), it is 
sufficient to prove existence of \(T\) in the latter case. 

Let \[
w(\alpha,\alpha')=\frac{\alpha(e^{\alpha'}-1)}{\alpha'(e^\alpha-1)};\]
we recall (from \cite{Dabrowski:2009hv}; but see 
from\footnote{Here, we consider 
the star product defined with the ``symmetric'' Weyl prescription for Weyl operators, not to be confused with the ``time first'' and ``space first'' 
prescriptions. The symmetric form of the Weyl operators was proposed in 
\cite{Agostini:2002de}, based on the integration of the BCH formula
of \cite{Kosinski:1999dw}. In connection with this, see also \cite{Dimitrijevic:2003wv}.}) that in \(1+1\) dimensions
\[
(\hat f\ast_0\hat g)(\alpha,\beta)
=\int  d\alpha'd\beta'w(\alpha-\alpha',\alpha)\hat f(\alpha',\beta')
\hat g(\alpha-\alpha',w(\alpha-\alpha',\alpha)\beta-
w(\alpha'-\alpha,\alpha')\beta').
\]
Then
\[
(T\;\hat f\otimes\hat g)(\alpha,\beta,\alpha',\beta')=
w(\alpha',\alpha'+\alpha)\hat f(\alpha,\beta)\hat g
(\alpha',(w(\alpha',\alpha'+\alpha)-w(\alpha',\alpha))\beta-
w(\alpha',\alpha)\beta')
\]
fulfils \eqref{eq:tw_four}, 
where \(\ast_0\) is now the deformed \(\kappa\)-Minkowski product 
in Fourier space for \(1+1\) dimensions. 
Moreover,
\[
\hat f(\alpha,\beta)\hat g(\alpha',\beta')=
w(\alpha'+\alpha,\alpha')
(T\;\hat f\otimes\hat g)(\alpha,\beta,\alpha',(1-w(\alpha'+\alpha,\alpha))\beta-\beta'),
\]
which gives invertibility of \(T\).

\section{Deformed Covariant $\kappa$-Minkowski as a Non Invariant Reduction }

Now, we come back to the fully covariant model. In this case
the presence of a non trivial centre forbids to define a star product
through a quantization prescription of the kind of \eqref{eq:weyl_quant} 
for symbols only depending on \(x\), since 
then a requirement of the form \eqref{eq:star_prod} would be inconsistent: 
indeed the right hand side would
fail to be an object of the form of the left hand side. To say it differently,
there is the necessity of considering a~\(v\)~dependence to account for the centre
of the algebra. This problem
has been thoroughly discussed in \cite{Doplicher:1994tu} in a different 
context (see also the less technical \cite{Doplicher:1994zv}), 
so that we only shortly describe the analogous solution in our case. 

We consider a more general class of symbols,
namely functions \(f=f(v,x)\) of \(H\times\mathbb R^4\), where the 
two--sheeted, mass 1  
hyperboloid \(H\)  arises as the set of common generalised eigenvalues (joint spectrum) of the pairwise commuting operators \(V^\mu\) (see the appendix). 
For such symbols, we define a quantization in two steps. First, for each
\(x\) fixed, we replace \(v\) by \(V\) in the usual sense
of functions of pairwise commuting operators, so to obtain
an operator valued function
\[
x\mapsto f(V,x).
\]
Next we consider the quantization
\[
f(V,X)=\frac{1}{(2\pi)^2}\int d\alpha\; \hat f(V,\alpha)e^{i\alpha_\mu X^\mu},
\]
where
\[
\hat f(V,\alpha)=\frac{1}{(2\pi)^2}\int dx\; 
\hat f(V,x)e^{-i\alpha_\mu x^\mu}.
\]
The above prescription is unambiguous, since \([X^\mu,V^\nu]=0\).
Now, 
\[
(f\star g)(V,X)=f(V,X)g(V,X)
\]
gives a consistent definition of the star product. 
The *-algebra of such generalised symbols is sufficient
to fully describe the operator algebra arising from quantization. 

Note that, for each 
\(v\in H\)
fixed, we may define a deformed product ``at \(v\)''  by
\[
f(v,\cdot)\star_v g(v,\cdot)=(f\star g)(v,\cdot),
\]
which is meaningful as a product of reduced symbols depending on \(x\) only.
Every such star product is precisely the star product which would be defined
by  
\[
f(v,X_{(v)})g(v,X_{(v)})=(f(v,\cdot)\star_v g(v,\cdot))(X_{(v)}),
\]
if \(X_{(v)}\) were the coordinates defined by \eqref{eq:v_Coord} of the 
appendix, and \(f(v,\cdot),g(v,\cdot)\) are thought of as functions
of \(x\) only, parametric in \(v\).

In particular, \(\star_{{{v_{(0)}}}}\) is precisely the star product \(\star_0\)
of the usual \(\kappa\)-Minkowski spacetime, introduced in the preceding 
section.

We may define an action of the Lorentz group on the algebra of
generalised symbols, by
\[
(\tau_\varLambda f)(v,x)=f(\varLambda^{-1}v,\varLambda^{-1}x).
\]
If \(X,V,U\) is a covariant representation fulfilling 
(\ref{eq:kappa_cov_rel},\ref{eq:kappa_cov_unit}), then we find
\[
(\tau_\varLambda f)(V,X)=U(\varLambda)f(V,X)U(\varLambda)^{-1}.
\]
In particular, 
\[
(\tau_\varLambda f)\star(\tau_\varLambda g)(v,x)=
\tau_\varLambda(f\star g)(v,x).
\]
Now, we observe that
\[
(\tau_\varLambda f)\star(\tau_\varLambda g)(v,x)=
f(\varLambda^{-1}v,\varLambda^{-1}\cdot)
\star_{v}g(\varLambda^{-1}v,\varLambda^{-1}\cdot),
\]
so that in particular if we take generalised symbols \(f,g\)  
which are constant in \(v\) for every \(x\),
and we evaluate  the above at
\(v=\varLambda{v_{(0)}}\), we get
\[
(\tau_\varLambda f)\star(\tau_\varLambda g)(\varLambda{v_{(0)}},x)=
((\alpha_\varLambda f)\star_{\varLambda{v_{(0)}}}(\alpha_\varLambda g))(x),
\]
where \(\alpha_\varLambda\) only affects the \(x\) dependence of symbols,
defined in the preceding section.
By comparison with \eqref{eq:def_cov} we recognise that 
\begin{equation}\label{eq:tw_cov_is_bad}
(\alpha_\varLambda f)\star_{\varLambda{v_{(0)}}}(\alpha_\varLambda g)=
m_{\star_0}\circ\tilde\alpha^{(2)}_\varLambda(f\otimes g); 
\end{equation}
so that the right hand side may be regarded merely
as an alternative notation for the left hand side.

In other words, 
deformed covariance (deformed Lorentz action, same product)
is equivalent to usual (form) covariance (usual Lorentz action, Lorentz 
transformed product), in complete analogy with the analysis of
\cite{Piacitelli:2009fa}.

Since this situation 
is not special to this model, but a consequence
of central covariantization which is always possible in principle,
in a sense even the existence of \(F\) 
is not essential, since in the end the only r\^ole it plays is that
of an equivalent notation for \(\star_{\varLambda{v_{(0)}}}\), which is always 
meaningful.
 
We now give a physical interpretation of the above formal developments.
Localisation states
on the algebra are linear functionals 
\[
\omega(f)=\int\limits_{H\times \mathbb R^4} dv\;dx\;
\rho(v,x)f(v,x);
\] 
it is rather difficult to express in terms of conditions on the kernel \(\rho\)
the fundamental properties of a state, in particular positivity:
\[
\omega(\bar f\star f)\geqslant 0.
\]
However we may select a special class of states by requiring that, 
when restricted to functions of \(v\) only, they are atomic measures 
concentrated on \({{v_{(0)}}}\), namely states of the form
\[
\omega(f)=\int dx\; \rho'(x)f({{v_{(0)}}},x)
\]
for some \(\rho'\).
If we took as a fundamental assumption of the theory that the only admissible
localisation states were precisely those appearing in the above form
to some given observer (conventionally called the privileged one), he 
only would ``see''
the usual \(\kappa\)-Minkowski spacetime, and all the rest of the structure 
would remain hidden to him. He would be naturally led to 
use symbols not depending on \(v\), and 
the product \(\star_0=\star_{{{v_{(0)}}}}\). 

Another observer, connected to the privileged one by \(\varLambda\),
only could access the states obtained
by the pull back action of the Lorentz group on symbols, so she only
could see symbols evaluated at \(v=\varLambda{v_{(0)}}\). She would use the
product \(\star_{v}\), which can be written equivalently using
deformed covariance.

Hence the formalism of deformed covariance is equivalent to 
work with a fully covariant model, up to arbitrarily dismissing a huge family
of otherwise admissible localisation states. It is this very last step which
destroys covariance.

\section{Conclusions}

Whenever a set of more or less 
physically motivated commutation relations is given, which
define a candidate for noncommutative spacetime, then full 
Lorentz symmetry is not an issue. We 
have seen that there is a standard strategy (covariantization)
leading to a fully covariant model. Of course, in principle there may be 
other~---~more complex~---~strategies which do not end up with central extensions. Anyhow,
whatever strategy is taken, if physical motivations really are motivating
one should check whether they survive the covariantization or not. In the
first case, all is well that ends well; 
in the second case, motivations might 
be so strong to make us consider the breakdown of the relativity of 
observers as an acceptable cost. Or they might not: 
in the absence of strong physical motivations 
a non covariant choice has no evident payoff.

We noted that the approach of deformed covariance (in the spirit of twisted 
covariance) is possible whenever a certain 
invertible deformation operator \(F\) exists; 
its exact form is irrelevant, existence is all one needs. Indeed, we might
go one step further: neither existence of \(F\) is necessary, 
\(F\) can be postulated. Then everything seems to go fine, since all equations 
can be given sense {\itshape a posteriori} in terms of the corresponding 
covariantization, even if the deformation operator eventually turns
out not to exist. 

In the end, the only requirement 
which is made on the deformation operator \(F\) 
is to reproduce the deformed product when
\(F f\!\otimes\! g\) is restricted to the diagonal. Hence \(F\) is highly non unique, and 
undetermined off the diagonal. Any physical consequence deriving from the off diagonal
behaviour of \(F\) would require some argument motivating 
the particular choice of \(F\). Otherwise, for any formal development which 
does not depend
on the particular choice of the off diagonal behaviour of \(F\), one is entitled to conjecture
that it can be reproduced in the covariantized setting as well.

This seems to suggest
that the deformation operator 
might contain no additional physically interesting  
information with respect to
the underlying covariant algebra. 

One substantial difference between the two approaches to covariance 
appears to be the possibility of considering 
commutation relations depending on the spacetime event, in connection with gauge theories.
This is sometimes 
claimed to only be possible in the formalism of deformed covariance,
at least in the case of twists. However, there are two 
aspects which should be thoroughly discussed. 

Indeed, (i) we should agree on the meaning of 
``possible''. If one takes existence of operators fulfilling 
regular commutation relations as fundamental, 
it is not clear that locally twisted commutation relations among the coordinates 
admit any representation by selfadjoint operators; this is not an idle problem, since
the representations determine the full algebraic content (the universal C*-algebra) of the theory.
On the contrary, if deformed products are taken as fundamental (namely disregarding
representations), the language of operators and their formal commutation relations 
should be considered as unavailable.

Last but not least, (ii) when speaking of local twists we implicitly
give a meaning to infinitely small
(classical) points of classical spacetime. 
This appears to be in plain contrast with the 
fundamental idea of noncommutative geometry, where the concept of ``point'' is dismissed as a fundamental one. 
Classical points may be given a meaning only as derived 
concepts; yet the family of 
classical points (which may well be empty as in the DFR model) 
can be sufficient to realise the whole spectrum of the 
localisation algebra in the commutative case only (Gel'fand's theorem).

\section*{Mathematical Appendix}
\label{sec:cov_mod}
We 
show here that there is a (essentially unique) covariant representation
of \eqref{eq:kappa_cov_rel}, namely selfadjoint 
operators \(X^\mu,V^\mu\)  and a strongly continuous unitary representation \(U\) of the Lorentz group, fulfilling 
(\ref{eq:kappa_cov_rel},\ref{eq:kappa_cov_unit}).

We first focus on irreducible
representations:
by Schur's lemma, all central quantities must be multiples
of the identity in an irreducible representation, so that we must have 
\(V^\mu=v^\mu I\)
for some real vector \(v\in H=H_-\cup H_+\), the mass 1 hyperboloid with 
connected components
\[
H_\pm=\{v\in\mathbb R^4:\pm v^0>0, v_\mu v^\mu=1\}.
\]
Here \(H\) plays the a r\^ole analogous to that of the manifold 
\(\Sigma\) in \cite{Doplicher:1994tu}.
In this case, we say we face an irreducible representation belonging to \(v\), for short.
Irreducible representations belonging to different vectors are clearly inequivalent.
Let now \(v,v'\) be any two such vectors; there always is some Lorentz matrix
\(\varLambda\) such that \(\varLambda v=v'\). Given an irreducible representation \(X\) 
belonging to \(v\),
\(X'=\varLambda X\) also is a representation, which belongs to \(v'\).
Since \(\varLambda\) is invertible, \(X'\) also
is irreducible. Taking linear combinations commute with 
the adjoint action of unitary operators; hence
this correspondence sends equivalence classes belonging to \(v\) 
into equivalence classes belonging to \(v'\), and vice versa. Thus, it 
suffices to classify irreducible representations belonging to 
one only \({{v_{(0)}}}\), e.g. 
\[
{{v_{(0)}}}=(1,0,0,0).
\]

We now may take profit 
from the fact \cite{Dabrowski:2010yk} that all irreducible 
representations belonging to \({{v_{(0)}}}\) 
appear in the disintegration of the following,
universal representation
\begin{gather*}
X_{({{v_{(0)}}})}^0=I\otimes P,\\
X_{({{v_{(0)}}})}^j=C^j\otimes e^{-Q}
\end{gather*}
acting on the Hilbert space \(\mathfrak H_0=\mathfrak K\otimes L^2(\mathbb R)\),
where \(P,Q\) are the usual Schr\"odinger operators on the line, fulfilling
\([P,Q]=-iI\); while the \(C^j\)s are pairwise commuting bounded
selfadjoint operators on some Hilbert space \(\mathfrak K\) with joint 
spectrum\footnote{We recall that if \(\psi\) is a common eigenvector
for the operators \(A_1,\dotsc,A_n\), so that \(A_j\psi=a_j\psi\),
then the \(n\)-tuple \((a_1,\dotsc,a_n)\) is in the joint spectrum 
of \(A_1,\dotsc,A_n\), which is a subset of \(\mathbb R^n\). 
The joint spectrum also contains all \(n\)-tuples 
corresponding to possibly generalised common eigenvectors, 
so that some \(a_j\) in a given \(n\)-tuple
may well belong to the continuous spectrum of the corresponding \(A_j\).
} 
\(S^2\cup \{0\}\subset\mathbb R^3\). It follows 
that \(E=\sum_j{C^j}^2\) is an orthogonal projection, such that
\(R^2:=\sum_j{X^j}^2=E\otimes e^{-2Q}\); hence this quantization only 
involves time and the distance from the origin, while angle variables
remain classical. In this sense the quantization is radial; it may be obtained 
equivalently from the relations
\begin{subequations}\label{eq:radial_rels}
\begin{gather}
[T,R]=iR,\label{eq:radial_rels_1dim}\\
[T,C_j]=[R,C_j]=0
\end{gather}
up to setting
\begin{equation}
X_j=C_jR;
\end{equation}
\end{subequations}
in other words 
the full relations define a central extension of the 1+1 relation 
\eqref{eq:radial_rels_1dim}. See
\cite{Dabrowski:2010yk} for a complete discussion.

Note also that \((I-E)\otimes I\) is a central orthogonal projection 
onto the subspace where all \(C^j\)'s vanish, as well as the 
\(X^j\)'s and \(R\). Indeed, among all possible representations of the 
\(\kappa\)-Minkowski relations there are the trivial ones, where 
all the space coordinates are zero;
omitting them would be equivalent to removing the time axis through the origin.

It follows that all possible equivalent representations belonging to any
\(v\in H\) can be obtained from the disintegration of 
\begin{equation}\label{eq:v_Coord}
{X_{(v)}}^\mu={\varLambda^{\mu}}_\nu X^\nu_{({{v_{(0)}}})}, 
\end{equation}
where \(\varLambda{v_{(0)}}=v\).

We are now ready to construct the covariant representation; instead of using
direct integrals, we simply describe the result:
let
\[
\mathfrak H=L^2(H,dv;\mathfrak H_0)
\]
be the space of \(L^2\) vector valued functions of \(H\), with values in 
\(\mathfrak H_0\), 
where \(dv\) is the Lorentz invariant measure on the mass 1 hyperboloid \(H\).
Moreover, let \(\varLambda(v)\) be a continuous function of \(H\)
with values in \(\mathscr L\), fulfilling \(\varLambda(v){v_{(0)}}=v\). Then set
\begin{align*}
(X^\mu\Psi)(v)&={\varLambda(v)^\mu}_\nu X^\nu_{({{v_{(0)}}})}\Psi(v),\\
(V^\mu\Psi)(v)&=v^\mu\Psi(v),\\
(U(\varLambda)\Psi)(v)&=\Psi(\varLambda^{-1}v),
\end{align*}
where \(v\in H,\varLambda\in\mathscr L\) and we recall that, 
for each \(v\), \(\Psi(v)\in \mathfrak H_0\).

By construction these operators fulfil 
(\ref{eq:kappa_cov_rel},\ref{eq:kappa_cov_unit}).
The above representation can be easily shown not to depend on the choice 
of the map \(\varLambda(v)\).

By standard techniques (see again \cite{Dabrowski:2010yk}), the universal 
C*-algebra generated by \(f(V,X)\)
(as \(f\) runs in the admissible symbols)
can be shown to be 
\[
\mathcal C_\infty(H)\otimes
\big((\mathcal C(S^2)\otimes\mathcal K)\oplus \mathcal C_\infty(\mathbb R)
\big),\]
where \(\mathcal C_\infty\) means continuous and vanishing at infinity.
The picture is as follows: 
the compact C*-algebra \(\mathcal K\) arises from the 
quantization of time and radius, the latter being strictly greater 
than zero; the commutative factor
\(\mathcal C(S^2)\) describes the classical angle variables.
Hence \(\mathcal C(S^2)\otimes\mathcal K\) describes the quantization
of the Minkowski spacetime with the time axis through the origin removed;
adding \(\mathcal C_\infty(\mathbb R)\) restores the missing axis as
a classical submanifold which is topologically disjoint from the rest.
Finally \(H\) appears as additional degrees of freedom, surviving the
classical limit as a hidden manifold (extra dimensions).

\footnotesize
\bibliographystyle{utphys}
\bibliography{mybibtex}

\end{document}